\documentstyle[prd,aps,preprint,tighten,epsfig]{revtex}

\begin{document}

\draft

\title{Model-independent Constraint on the Neutrino Mass Spectrum \\
from the Neutrinoless Double Beta Decay}
\author{\bf Zhi-zhong Xing}
\address{Institute of High Energy Physics, P.O. Box 918 (4), 
Beijing 100039, China \\
({\it Electronic address: xingzz@mail.ihep.ac.cn}) }
\maketitle

\begin{abstract}
We present a concise formula to relate the effective mass term of
the neutrinoless double beta decay to a single neutrino mass,
two Majorana CP-violating phases and four observables of neutrino 
oscillations for a generic neutrino mass spectrum. If the alleged 
evidence for the neutrinoless double beta decay is taken into account, 
one may obtain a rough but model-independent constraint on the absolute 
scale of neutrino masses -- it is most likely to be
in the range between 0.1 eV and 1 eV. 
\end{abstract}

\pacs{PACS number(s): 14.60.Pq, 13.10.+q, 25.30.Pt} 

\newpage

\section{Introduction}

The solar and atmospheric neutrino oscillations observed 
in the Super-Kamiokande experiment \cite{SK} have
provided convincing evidence that neutrinos are massive and lepton
flavors are mixed. If neutrinos are Majorana particles, 
a complete description of the flavor mixing phenomenon in the framework 
of three lepton families requires six real parameters: 
three mixing angles, one Dirac-type CP-violating phase 
and two Majorana-type CP-violating phases. So far some preliminary 
knowledge on three flavor mixing angles and two neutrino mass-squared
differences have been achieved from current neutrino oscillation 
experiments. It is likely to determine the Dirac-type CP-violating phase 
from a new generation of accelerator neutrino 
experiments with very long baselines, if the solar neutrino anomaly is
attributed to the large-angle Mikheyev-Smirnov-Wolfenstein (MSW) 
oscillation \cite{MSW} and the flavor mixing angle
between the first and third lepton families is not too small. 
To pin down two Majorana phases is practically impossible, however, since 
all possible lepton-number-nonconserving processes induced by light
Majorana neutrinos are suppressed in magnitude by extremely small
factors compared to normal weak interactions \cite{FX01}. 
The only experimental possibility to get some information on two 
Majorana-type CP-violating phases is to measure the neutrinoless double 
beta decay.

Recently Klapdor-Kleingrothaus {\it et al} have reported their first
evidence for the existence of the neutrinoless double beta decay \cite{K}. 
At the $95\%$ confidence level, the effective mass term of the neutrinoless
double beta decay is found to lie in the following range:
\begin{equation}
0.05 ~ {\rm eV} \; \leq \; \langle m\rangle_{ee} \; \leq \; 0.84 ~ {\rm eV}
\; .
\end{equation}
A number of authors have discussed the implications of this alleged
evidence on neutrino masses \cite{Glashow} and textures of the
neutrino mass matrix \cite{Ma,KS1,Uehara,Vissani,Hambye,KS2}.

The purpose of this paper is two-fold. First, we present a concise
formula to relate $\langle m\rangle_{ee}$ to a single neutrino mass,
two Majorana phases and four observables of neutrino oscillations
for a generic neutrino mass spectrum. Second, we take the experimental 
result in Eq. (1) seriously and
obtain a rough but model-independent constraint on the absolute
scale of neutrino masses -- it is most likely to be 
in the range between 0.1 eV and 1 eV. This result implies that three 
neutrino masses are nearly degenerate.

\section{Formulation}

Current experimental data \cite{SK,SNO} 
indicate that solar and atmospheric neutrino 
oscillations are dominated by $\nu_e \rightarrow \nu_\mu$ and
$\nu_\mu \rightarrow \nu_\tau$ transitions, respectively. The
neutrino mass-squared differences associated with solar and atmospheric
neutrino oscillations are thus defined as
\begin{eqnarray}
\Delta m^2_{\rm sun} & \equiv & \left | m^2_2 ~ - ~ m^2_1 \right | \; ,
\nonumber \\
\Delta m^2_{\rm atm} & \equiv & \left | m^2_3 ~ - ~ m^2_2 \right | \; ,
\end{eqnarray}
where $m_i$ (for $i=1,2,3$) denote the mass eigenvalues of three
neutrinos. Without loss of generality, we require $m_i$ to be real and
positive. The observed hierarchy between
$\Delta m^2_{\rm sun}$ and $\Delta m^2_{\rm atm}$ can tell the
relative sizes of three neutrino masses, but it cannot shed any light
on the absolute value of $m_1$, $m_2$ or $m_3$. In order to show how the 
absolute scale of neutrino masses can be constrained from the 
neutrinoless double beta decay, we express $m_1$ and $m_2$ in terms of
$m_3$, $\Delta m^2_{\rm sun}$ and $\Delta m^2_{\rm atm}$ with the help
of Eq. (2) \cite{Beta}. The results are concisely summarized as 
\begin{eqnarray}
m_1 & = & \sqrt{m^2_3 ~ + ~ p \Delta m^2_{\rm atm} ~ + ~
q \Delta m^2_{\rm sun}} \;\; ,
\nonumber \\
m_2 & = & \sqrt{m^2_3 ~ + ~ p \Delta m^2_{\rm atm}} \;\; ,
\end{eqnarray}
where $p=\pm 1$ and $q=\pm 1$ stand for four possible patterns of the
neutrino mass spectrum:
$$
(p, q) = (+1, +1): ~~ m_1 > m_2 > m_3 \; ;
\eqno{\rm (4a)}
$$
$$
(p, q) = (-1, -1): ~~ m_1 < m_2 < m_3 \; ;
\eqno{\rm (4b)} 
$$
$$
(p, q) = (+1, -1): ~~ m_1 < m_2 > m_3 \; ;
\eqno{\rm (4c)}
$$
$$
(p, q) = (-1, +1): ~~ m_1 > m_2 < m_3 \; .
\eqno{\rm (4d)}
$$
We see that $m_1 \approx m_2$ holds as a straightforward consequence
of $\Delta m^2_{\rm sun} \ll \Delta m^2_{\rm atm}$. The signs of
$p$ and $q$ can be determined from the future long-baseline 
neutrino oscillation experiments with high-quality conventional
neutrino beams \cite{Hagiwara} or at neutrino factories \cite{Factory}. 

As solar and atmospheric neutrino oscillations are approximately
decoupled from each other, their mixing factors $\sin^2 2\theta_{\rm sun}$
and $\sin^2 2\theta_{\rm atm}$ may have simple relations with the
matrix elements of the lepton flavor mixing matrix $V$, which is 
defined to link the neutrino mass eigenstates $(\nu_1, \nu_2, \nu_3)$ to 
the neutrino flavor eigenstates $(\nu_e, \nu_\mu, \nu_\tau)$:
\setcounter{equation}{4}
\begin{equation}
\left ( \matrix{
\nu_e \cr
\nu_\mu \cr
\nu_\tau \cr} \right ) =
\left ( \matrix{
V_{e1}		& V_{e2}	& V_{e3} \cr
V_{\mu 1}	& V_{\mu 2}	& V_{\mu 3} \cr
V_{\tau 1}	& V_{\tau 2} 	& V_{\tau 3} \cr} \right )
\left ( \matrix{
\nu_1 \cr
\nu_2 \cr
\nu_3 \cr} \right ) \; .
\end{equation}
The mixing factor associated with the CHOOZ (or Palo Verde) reactor 
neutrino oscillation experiment \cite{CHOOZ}, 
denoted as $\sin^2 2\theta_{\rm chz}$, 
is also a simple function of the matrix elements of $V$ in the same 
approximation. The explicit expressions of $\sin^2 2\theta_{\rm sun}$,
$\sin^2 2\theta_{\rm atm}$ and $\sin^2 2\theta_{\rm chz}$ read as
follows:
\begin{eqnarray}
\sin^2 2\theta_{\rm sun} & = & 4 |V_{e1}|^2 |V_{e2}|^2 \; ,
\nonumber \\
\sin^2 2\theta_{\rm atm} & = & 4 |V_{\mu 3}|^2 
\left ( 1 - |V_{\mu 3}|^2 \right ) \; ,
\nonumber \\ 
\sin^2 2\theta_{\rm chz} & = & 4 |V_{e3}|^2 
\left ( 1 - |V_{e3}|^2 \right ) \; .
\end{eqnarray}
Taking the unitarity of $V$ into account, one may reversely express 
$|V_{e1}|^2$, $|V_{e2}|^2$, $|V_{e3}|^2$ and $|V_{\mu 3}|^2$ in terms
of $\theta_{\rm sun}$, $\theta_{\rm atm}$ and $\theta_{\rm chz}$:
\begin{eqnarray}
|V_{e1}|^2 & = & \frac{1}{2} \left ( \cos^2\theta_{\rm chz}
+ \sqrt{\cos^4\theta_{\rm chz} - \sin^2 2\theta_{\rm sun}} \right ) \; ,
\nonumber \\
|V_{e2}|^2 & = & \frac{1}{2} \left ( \cos^2\theta_{\rm chz}
- \sqrt{\cos^4\theta_{\rm chz} - \sin^2 2\theta_{\rm sun}} \right ) \; ,
\nonumber \\
|V_{e3}|^2 & = & \sin^2 \theta_{\rm chz} \; ,
\nonumber \\
|V_{\mu 3}|^2 & = & \sin^2 \theta_{\rm atm} \; .
\end{eqnarray}
Current experimental data favor $\theta_{\rm chz} \ll 1$ and
$\theta_{\rm sum} \sim \theta_{\rm atm} \sim 1$, therefore 
$|V_{e1}|^2 \sim |V_{e2}|^2 \sim |V_{\mu 3}|^2 \gg |V_{e3}|^2$
is expected to hold.

Note that only the matrix elements $V_{e1}$, $V_{e2}$ and
$V_{e3}$ are relevant to the neutrinoless double beta decay. Without loss
of generality, one may redefine the phases of three charged lepton
fields in an appropriate way such that the phases of
$V_{e1}$ and $V_{e2}$ are purely of the Majorana type and 
$V_{e3}$ is real \cite{FGM}. In other words,
\begin{equation}
\arg (V_{e1}) = \rho \; , ~~ \arg (V_{e2}) = \sigma \; ,
~~ \arg (V_{e3}) = 0 \; .
\end{equation}
Of course $\rho$ and $\sigma$ do not have any effect on CP or T
violation in normal neutrino-neutrino and antineutrino-antineutrino 
oscillations \cite{Xing00}. With the help of Eqs. (3),
(7) and (8), we then arrive at a model-independent expression for
the effective mass term of the neutrinoless double beta decay:
\small
\begin{eqnarray}
\langle m \rangle_{ee} & = & \left | m_1 V^2_{e1} ~ + ~
m_2 V^2_{e2} ~ + ~ m_3 V^2_{e3} \right | 
\nonumber \\ 
& = & \left | m_3 \sin^2 \theta_{\rm chz} ~ 
+ ~ \frac{\cos^2\theta_{\rm chz}}{2} 
\left ( \sqrt{m^2_3 + p \Delta m^2_{\rm atm} + q \Delta m^2_{\rm sun}}
~ e^{2i\rho} + \sqrt{m^2_3 + p \Delta m^2_{\rm atm}} ~ e^{2i\sigma} \right )
\right . 
\nonumber \\
&& \left . + \frac{\sqrt{\cos^4\theta_{\rm chz} - \sin^2 2\theta_{\rm sun}}}
{2}\left (\sqrt{m^2_3 + p \Delta m^2_{\rm atm} + q \Delta m^2_{\rm sun}}
~ e^{2i\rho} - \sqrt{m^2_3 + p \Delta m^2_{\rm atm}} ~ e^{2i\sigma} \right )
\right | \; .
\end{eqnarray}
\normalsize
One can see that $\langle m\rangle_{ee}$ consists of three unknown
parameters: $m_3$, $\rho$ and $\sigma$, which are unable to be determined
from any neutrino oscillation experiments. Once $\Delta m^2_{\rm sun}$,
$\Delta m^2_{\rm atm}$, $\theta_{\rm sun}$, $\theta_{\rm atm}$ and
$\theta_{\rm chz}$ are measured to an acceptable degree of accuracy, we
will be able to get a useful constraint on the absolute neutrino mass $m_3$ 
for arbitrary values of $\rho$ and $\sigma$ from the observation of
$\langle m\rangle_{ee}$. If the magnitude of $m_3$ could roughly be known 
from some cosmological constraints, it would be likely to obtain some 
loose but instructive information on the Majorana phases $\rho$ and
$\sigma$ by confronting Eq. (9) with the experimental result of 
$\langle m\rangle_{ee}$. Anyway further progress in our
theoretical understanding of the origin of neutrino masses and CP
violation is crucial for a complete determination
of the free parameters under discussion.

\section{Illustration}

Now let us illustrate the dependence of $\langle m\rangle_{ee}$ on $m_3$,
$\rho$ and $\sigma$ numerically. 
Assuming that the solar neutrino anomaly is attributed
to the large-angle MSW effect \cite{MSW},
we typically take $\Delta m^2_{\rm sun} = 5\cdot 10^{-5} ~ {\rm eV}^2$
and $\sin^2 2\theta_{\rm sun} = 0.8$. We choose 
$\Delta m^2_{\rm atm} = 3\cdot 10^{-3} ~ {\rm eV}^2$ and
$\sin^2 2\theta_{\rm atm} =1$ for the atmospheric neutrino oscillation.
In addition, we use the typical value $\sin^2 2\theta_{\rm chz} =0.05$
in our numerical calculations, which is consistent with the upper bound
$\sin^2 2\theta_{\rm chz} <0.1$ from the CHOOZ reactor neutrino 
experiment \cite{CHOOZ}
\footnote{Note that $\sin^2 2\theta_{\rm chz} \sim 0.05$ is also favored 
in a number of phenomenological models of lepton mass matrices. 
See Ref. \cite{Review} for a review with extensive references.}.
The Majorana phases $\rho$ and $\sigma$ are completely unknown. To 
illustrate, we consider four instructive possibilities for $\rho$ and 
$\sigma$: (1) $\rho = \sigma = 0$; (2) $\rho = \pi/4$ and $\sigma = 0$;
(3) $\rho = 0$ and $\sigma = \pi/4$; and (4)
$\rho =\sigma = \pi/4$. Our results for $\langle m\rangle_{ee}$ as a function
of $m_3$ are shown in FIG. 1, where all possible patterns of 
the neutrino mass spectrum as listed in Eq. (4) have been taken into account. 
Some comments are in order.

(1) A careful analysis shows that
the result of $\langle m\rangle_{ee}$ is essentially insensitive to
the sign of $q$. In other words, the cases $m_1 > m_2$ and $m_1 < m_2$ 
are almost indistinguishable in the neutrinoless double beta decay. 
This feature is a straightforward consequence of the hierarchy
$\Delta m^2_{\rm sun} \ll \Delta m^2_{\rm atm}$. As shown in Eq. (9),
the contribution of $q\Delta m^2_{\rm sun}$ to $\langle m\rangle_{ee}$
is negligible unless a complete cancellation between $m^2_3$ and
$p\Delta m^2_{\rm atm}$ terms happens to take place.

(2) When $p =-1$ (i.e., $m_2 < m_3$), Eq. (3) implies that
$m_3$ has the following lower bound 
\begin{equation}
m_3 \; \geq \; \left \{ \matrix{ 
\sqrt{\Delta m^2_{\rm atm}} ~~~~~~ (q=+1) \; , ~~~~~~~~~~~~ \cr\cr
\sqrt{\Delta m^2_{\rm atm} + \Delta m^2_{\rm sun}} ~~~~~~ (q=-1) \; . } 
\right .
\end{equation}
In view of the typical inputs of $\Delta m^2_{\rm atm}$ and 
$\Delta m^2_{\rm sun}$ taken above, we obtain 
$m_3 \geq 0.0548$ eV for $q=+1$ and $m_3 \geq 0.0552$ eV for $q=-1$.
Such lower bounds of $m_3$ have indeed been reflected in FIG. 1.

(3) We find that it is numerically difficult to distinguish between the 
possibilities $\rho =\sigma =0$ and $\rho =\sigma =\pi/4$. 
The reason is simply that the second term on the right-hand side of
Eq. (9) dominates over the other two terms, when $\rho =\sigma$ is taken.
Hence the value of $\langle m\rangle_{ee}$ becomes insensitive to the 
explicit values of two identical Majorana phases. One can also see that
the possibility of $\rho =0$ and $\sigma =\pi/4$ is almost indistinguishable
from the possibility of $\rho =\pi/4$ and $\sigma =0$. In this specific case
the first term on the right-hand side of Eq. (9) plays an insignificant role,
therefore the magnitude of $\langle m\rangle_{ee}$ is essentially invariant 
under an exchange of the values between $\rho$ and $\sigma$.

(4) We observe that the changes of $\langle m\rangle_{ee}$ are rather
mild for the typical values of $\rho$ and $\sigma$ chosen above. 
If reasonable inputs of $\sin^2 2\theta_{\rm sun}$, $\sin^2 2\theta_{\rm atm}$
and $\sin^2 2\theta_{\rm chz}$ are taken
\footnote{Only in a few extreme cases (e.g., 
$\cos^4 \theta_{\rm chz} \approx \sin^2 2\theta_{\rm sun}$ and 
$\rho \approx -\sigma$), which seem quite unlikely, large cancellations 
may take place in $\langle m\rangle_{ee}$ and make its magnitude 
significantly suppressed.},
a careful numerical scan shows that the magnitude of $\langle m\rangle_{ee}$
does not undergo any dramatic changes for arbitrary $\rho$ and $\sigma$. 
Thus a rough but model-independent
constraint on the absolute scale of neutrino masses can be obtained from
the observation of $\langle m\rangle_{ee}$. In view of the alleged 
experimental region
of $\langle m\rangle_{ee}$ in Eq. (1), we find that $m_3$ is most likely to
lie in the range 0.1 eV $\leq m_3 \leq$ 1 eV (see FIG. 1). 
This result is irrelevant to the details of four possible patterns of 
the neutrino mass spectrum. Note that $m_3 \geq 0.1$ eV 
implies that both $m_1 \approx m_3$ and $m_2 \approx m_3$ hold, as one
can see from Eq. (3). Therefore three neutrino masses
are nearly degenerate. Taking $m_3 = 0.5$ eV 
for example, we obtain $m_1 + m_2 +m_3 \approx 3 m_3 \approx 1.5$ eV. 
Such a sum of three neutrino masses can be translated in cosmology 
to $\Omega_\nu h^2 \approx 0.016$, where
$\Omega_\nu$ is the fraction of the critical density contributed by 
neutrinos and $h$ is the dimensionless Hubble constant. This typical
result is consistent with $\Omega_\nu h^2 \approx 0.05$ \cite{Wang},
extracted from the CMB measurements and galaxy cluster surveys.

\section{Conclusion}

We have presented a concise formula to relate the effective mass term of
the neutrinoless double beta decay to a single neutrino mass, two Majorana
phases and four observables of neutrino oscillations for four possible 
patterns of the neutrino mass spectrum. Taking into account the alleged 
evidence for the neutrinoless double beta decay, we have obtained a
rough but model-independent constraint on the absolute scale of active
neutrino masses: 0.1 eV $\leq m_3 \leq$ 1 eV. 
This result implies that three neutrino masses are nearly degenerate.
 
If the existence of the neutrinoless double beta decay can be confirmed,
it will be desirable to build new phenomenological models of lepton
mass matrices which can accommodate three neutrinos of nearly 
degenerate masses and reflect their Majorana nature. Further studies
of various lepton-number-violating processes will also become more
realistic and important.

\vspace{0.3cm}

\acknowledgments{
The author is grateful to H. Fritzsch and X.M. Zhang for stimulating
discussions. This work was supported in part by National Natural Science 
Foundation of China.

{\it Note added}: When this paper was being completed, the preprint of
Minakata and Sugiyama \cite{Minakata} appeared. Their best-fit analysis
leads to 0.11 eV $\leq \langle m\rangle_\beta \leq$ 1.3 eV for 
the mass parameter in the single beta decay experiments. This result
is consistent with ours for the absolute scale of neutrino masses.}

\newpage

\begin{figure}
\vspace{-0.2cm}
\epsfig{file=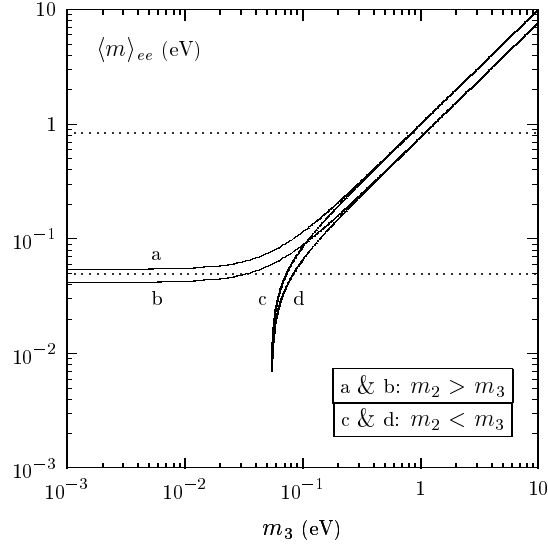,bbllx=1cm,bblly=4cm,bburx=20cm,bbury=32cm,%
width=15cm,height=22cm,angle=0,clip=}
\vspace{-10.3cm}
\caption{Illustrative dependence of $\langle m\rangle_{ee}$ on $m_3$ for the
neutrino mass spectrum $m_2 > m_3$ (curves a and b) and the
neutrino mass spectrum 
$m_2 < m_3$ (curves c and d), where we have typically taken
$\{\rho, \sigma \} = \{0, 0 \}$ or $\{ \pi/4, \pi/4 \}$ (curves a and c) 
and $\{\rho, \sigma \} = \{ 0, \pi/4 \}$ or $\{ \pi/4, 0 \}$ 
(curves b and d). The region between two dashed lines corresponds 
to the experimentally allowed values of $\langle m\rangle_{ee}$ at the $95\%$ 
confidence level.}
\end{figure}


\newpage
\begin{thebibliography}{99}

\bibitem{SK} Y. Fukuda {\it et al.}, Phys. Rev. Lett. {\bf 81}, 1562
(1998); {\it ibid.} {\bf 81}, 4279 (1998);
http://www-sk.icrr.u-tokyo.ac.jp/dpc/sk/.

\bibitem{MSW} S.P. Mikheyev and A. Yu Smirnov, Yad. Fiz. 
(Sov. J. Nucl. Phys.) {\bf 42}, 1441 (1985); 
L. Wolfenstein, Phys. Rev. D {\bf 17}, 2369 (1978).

\bibitem{FX01} J. Schechter and J.W.F. Valle,
Phys. Rev. D {\bf 23}, 1666 (1981); 
H. Fritzsch and Z.Z. Xing, Phys. Lett. B {\bf 517}, 363 (2001).

\bibitem{K} H.V. Klapdor-Kleingrothaus, A. Dietz, H.L. Harney,
and I.V. Krivosheina, Mod. Phys. Lett. A {\bf 16}, 2409 (2002).

\bibitem{Glashow} V. Barger, S.L. Glashow, D. Marfatia, and K. Whisnant,
hep-ph/0201262.

\bibitem{Ma} E. Ma, hep-ph/0201225.

\bibitem{KS1} H.V. Klapdor-Kleingrothaus and U. Sarkar,
hep-ph/0201226.

\bibitem{Uehara} Y. Uehara, hep-ph/0201277.

\bibitem{Vissani} F. Feruglio, A. Strumia, and F. Vissani, hep-ph/0201291.

\bibitem{Hambye} T. Hambye, hep-ph/0201307.

\bibitem{KS2} H.V. Klapdor-Kleingrothaus and U. Sarkar,
hep-ph/0202006.

\bibitem{SNO} SNO Collaboration, Q.R. Ahmad {\it et al.},
Phys. Rev. Lett. {\bf 87}, 071301 (2001).

\bibitem{Beta} F. Vassani, JHEP {\bf 9906}, 022 (1999);
S.M. Bilenkii, C. Giunti, W. Grimus,
B. Kayser, and S.T. Petcov, Phys. Lett. B {\bf 465}, 193 (1999); 
H.V. Klapdor-Kleingrothaus, H. P$\rm\ddot{a}$s,
and A.Yu. Smirnov, Phys. Rev. D {\bf 63}, 073005 (2001); 
S.M. Bilenky, S. Pascoli, and S.T. Petcov,
Phys. Rev. D {\bf 64}, 053010 (2001); 
Z.Z. Xing, Phys. Rev. D {\bf 64}, 093013 (2001).

\bibitem{Hagiwara} M. Aoki {\it et al.}, hep-ph/0112338;
and references therein.

\bibitem{Factory} B. Autin {\it et al.}, CERN 99-02 (1999); 
D. Ayres {\it et al.}, physics/9911009; 
C. Albright {\it et al.}, hep-ex/0008064; and references therein.

\bibitem{CHOOZ} CHOOZ Collaboration, M. Apollonio {\it et al.},
Phys. Lett. B {\bf 420}, 397 (1998); 
Palo Verde Collaboration, F. Boehm {\it et al.},
Phys. Rev. Lett. {\bf 84}, 3764 (2000).

\bibitem{FGM} P.H. Frampton, S.L. Glashow, and D. Marfatia,
hep-ph/0201008; 
Z.Z. Xing, hep-ph/0201151 (accepted for publication in Phys. Lett. B).

\bibitem{Xing00} Z.Z. Xing, Phys. Lett. B {\bf 487}, 327 (2000);
Phys. Rev. D {\bf 64}, 073014 (2001).

\bibitem{Review} H. Fritzsch and Z.Z. Xing, 
Prog. Part. Nucl. Phys. {\bf 45}, 1 (2000).

\bibitem{Wang} X. Wang, M. Tegmark, and M. Zaldarriaga,
astro-ph/0105091.

\bibitem{Minakata} H. Minakata and H. Sugiyama, hep-ph/0202003.

\end{thebibliography}
\end{document}